\documentclass[twocolumn]{aastex701}

\newcommand{\qth}{q_{\mathrm{th}}}

\usepackage{amsmath, amssymb}
\usepackage{cases}

\shorttitle{Mapping Giant Planet Occurrence with Stellar Mass and Orbital Separation}
\shortauthors{Chachan \& Fortney}

\begin{document}

\title{Mapping Giant Planet Occurrence with Stellar Mass and Orbital Separation: Implications for Gaia}

\author[0000-0003-1728-8269]{Yayaati Chachan}
\affiliation{Department of Astronomy and Astrophysics, University of California, Santa Cruz, CA95064, USA}
\email[show]{ychachan@ucsc.edu}  

\author[0000-0002-9843-4354]{Jonathan J. Fortney}
\affiliation{Department of Astronomy and Astrophysics, University of California, Santa Cruz, CA95064, USA}
\email{jfortney@ucsc.edu}  

\begin{abstract}
Giant planet demographics provide profound insights into their provenance. Although observations have established that giant planets are most common at 1-10 au around FGK stars and are rarer around M stars, much about the giant planet occurrence rate remains unknown. Here, we provide an important theoretical update to the occurrence rate function by studying how the planet mass distribution depends on stellar mass $M_\star$ and orbital separation $r$. We find that the maximum giant planet mass set by runaway accretion is $\approx 7.7~M_{\rm Jup} (M_\star / M_\odot)^{209/140} (r/1{\rm ~au})^{9/140} (t_{\rm d} / 3~{\rm Myr})^{3/7}$  and $\approx 5.5~M_{\rm Jup} (M_\star / M_\odot)^{95/98} (r/1{\rm ~au})^{18/49} (t_{\rm d} / 3~{\rm Myr})^{3/7}$ in viscously and passively heated regions, respectively, where $t_{\rm d}$ is the exponential decay timescale of the disk density. These estimates are obtained from analytical solutions for planet growth in the runaway accretion regime that provide an excellent match to numerical solutions. We incorporate these physically motivated constraints into the occurrence rate function of giant planets and show that the planet yield of Gaia DR4 is likely to be $\sim 5$ times smaller than expected for the case where planet mass distribution is independent of $M_\star$. In addition, we show that the orbital separation at which giant planet occurrence rate peaks would be $\propto M_\star$ if it is set by the water snowline and this would boost Gaia detections of $< 6~M_{\rm Jup}$ planets around K and M stars. 

\end{abstract}

\section{Introduction}

Observational demographics of giant planets are extremely informative for planet formation theories – since a large part of these theories revolve around creating these planetary behemoths in the first place. In the core accretion model, giant planets require the formation of a massive core that can accrete a substantial amount of H-He before the protoplanetary disk disperses \citep{Mizuno1980, Stevenson1982, Pollack1996a}. These cores must accrete enough gas to trigger runaway accretion during which gas cooling is rapid and planets transform into Jupiters by accreting hundreds of Earth masses of H-He gas. This picture makes several predictions, many of which have been borne out by observations. Giant planets are preferentially found around metal-rich stars \citep{Santos2001, Fischer2005}, in agreement with the importance of forming massive cores for giant planet formation. They also become more common with increasing stellar mass \citep{Johnson2010, Clanton2014, Fulton2021}, which combined with the observation that protoplanetary disks are more massive around higher mass stars \citep{Pascucci2016, Ansdell2017, Manara22}, suggests that massive disks are needed to spawn giant planets \citep{Bitsch2019, Schlecker2021, Chachan2022, Chachan2023a}. 

Given the dependence of giant planet occurrence rate and protoplanetary disk mass on stellar mass, it is important to understand how final planet mass varies with stellar mass. The final mass of a giant planet is set by runaway accretion, which must halt eventually due to gap opening and the dispersion of the protoplanetary disk \citep{Ginzburg2019, Rosenthal2020}. Theoretically understanding this relation between planet and stellar mass will have important bearing on interpreting existing observations \citep{VanZandt2026} and highlight future observations that can be used to test formation models. 

Giant planet occurrence rate is also a strong function of orbital separation. Long-term radial velocity surveys have revealed that giant planets seem to be most common at $1-10$ au around FGK stars \citep{Fernandes2019, Wittenmyer2019, Fulton2021}. This observation provides an important insight into the interplay of different processes that likely drive the existence of this peak: formation of more massive cores further out in the disk, dynamical times that enable assembly of planets within the disk lifetime, faster gas accretion in these regions, and the potentially paramount role of the water snowline for giant planet formation. Which of these processes is most important and how does it affect the peak location's variation with stellar mass? In turn, how do the final planet masses depend on orbital separation? 

These questions are particularly timely due to the imminent release of Gaia DR4 data, the upcoming Roman microlensing survey that will discover planets at $1-10$ au, and the Gaia DR5 data release in the future. They pose important consequences for expected planet yields from these missions and hold promise in revealing how giant planet masses and occurrence rate peak vary with stellar mass. Here, we present a theoretical prediction for how we expect final giant planet masses to depend on stellar mass and orbital separation. We also show that regions where massive cores form (by pebble or planetesimal accretion) and the water snowline location trend differently with stellar mass, thus making differing predictions for how the occurrence rate peak varies with stellar mass. Finally, we incorporate our physically motivated results into planet occurrence rate models to show their effect on the planet yield of Gaia DR4 and the properties of planets that we expect to find.

\section{Theory}

\subsection{Disk structure}

We are interested in the region of a protoplanetary disk where giant planets form and orbital distances at which Gaia will be able to detect giant planets. These regions, typically beyond $\sim 1$ au, are passively heated by stellar irradiation for lower stellar masses and viscously heated for higher stellar masses. We use the subscripts $_{\rm vis}$ and $_{\rm irr}$ to denote regions that are viscously and passively heated, respectively. For viscously heated regions, the disk temperature is given by \citep{Ida2016}
\begin{multline}
    T_{\rm vis} \simeq 255 \, {\rm K} \bigg(\frac{M_\star}{M_{\odot}}\bigg)^{3/10} \bigg(\frac{\dot{M}_\star}{10^{-8} M_{\odot} {\rm yr}^{-1}}\bigg)^{2/5} \\ \bigg(\frac{\alpha_{\rm t}}{3 \cdot 10^{-4}}\bigg)^{-1/5} \bigg(\frac{r}{1 \, {\rm au}}\bigg)^{-9/10}, 
    \label{eq:T_vis}
\end{multline}
where $r$ is the orbital distance and $M_\star$ is the stellar mass. The Shakura-Sunyaev parameter $\alpha_{\rm t}$ sets the strength of the turbulence and we adopt a fiducial value of $3 \cdot 10^{-4}$ in our work, which is intermediate in log-space between the accepted range of $10^{- 4} - 10^{-3}$ for this parameter. We use observations of stellar accretion rates to set $\dot{M}_\star$ and its dependence on $M_\star$ \citep{hartmannAccretionPreMainSequenceStars2016}:
\begin{equation}
    \dot{M}_\star = f_{\dot{M}} 10^{-8} (M_\star / M_\odot)^2 \, M_\odot \, {\rm yr}^{-1}. 
\end{equation}
Here, $f_{\dot{M}}$ controls the baseline accretion rate for a solar mass star and it can vary between $\sim 0.1 - 10$. In this study, we fix $f_{\dot{M}} = 1$ and drop this term from equations going forward. The expression for viscous heating then simplifies as follows:
\begin{equation}
    T_{\rm vis} \simeq 255 \, {\rm K} ~ \bigg(\frac{M_\star}{M_{\odot}}\bigg)^{11/10} \bigg(\frac{\alpha_{\rm t}}{3 \cdot 10^{-4}}\bigg)^{-1/5} \bigg(\frac{r}{1 \, {\rm au}}\bigg)^{-9/10}.
    \label{eq:T_vis_simplified}
\end{equation}
As the contribution of viscous heating ebbs, the temperature of the disk is set by incident radiation from the host star with luminosity $L_\star$ and is given by \citep{Chiang1997, Ida2016}
\begin{equation}
    T_{\rm irr} \simeq 150 \, {\rm K} \bigg(\frac{L_\star}{L_{\odot}}\bigg)^{2/7} \bigg(\frac{M_\star}{M_{\odot}}\bigg)^{-1/7} \bigg(\frac{r}{1 \, {\rm au}}\bigg)^{-3/7}.
    \label{eq:T_irr}
\end{equation}
During the protoplanetary disk phase, the stellar luminosity can be approximated to vary as $L_\star \propto M_\star^{3/2}$ \citep{Chachan2023a}, simplifying the temperature's dependence on $M_\star$ as
\begin{equation}
     T_{\rm irr} \simeq 150 \, {\rm K} \bigg(\frac{M_\star}{M_{\odot}}\bigg)^{2/7} \bigg(\frac{r}{1 \, {\rm au}}\bigg)^{-3/7}.
     \label{eq:T_irr_simplified}
\end{equation}
The temperature at a given location is then obtained by min(2000 K, max$(T_{\rm vis}, T_{\rm irr})$), where we cap the maximum temperature at 2000 K. The disk transitions from being viscously heated to being passively heated at orbital distance $r_{\rm vis \rightarrow irr}$
\begin{equation}
    r_{\rm vis \rightarrow irr} \simeq 3 ~{\rm au} ~ \bigg(\frac{M_\star}{M_{\odot}}\bigg)^{19/11} \bigg(\frac{\alpha_{\rm t}}{3 \cdot 10^{-4}}\bigg)^{-14/33}.
    \label{eq:r_vis_irr}
\end{equation}
The disk gas surface density is calculated assuming a steady-state viscously accreting disk with accretion rate $\dot{M}_\star$:
\begin{equation}
    \Sigma_{\rm g} = \frac{\dot{M}_\star}{3 \pi \nu}.
\end{equation}
Here, $\nu = \alpha_{\rm t} c_s H_{\rm g}$ is the kinematic viscosity of the disk, $c_s = \sqrt{k_B T / \mu m_{\rm H}}$ is the isothermal sound speed of the disk gas (with mean molecular weight $\mu = 2.3$), $H_g = c_s / \Omega$ is the disk scale height, and $\Omega$ is the Keplerian frequency. Using the temperature structure and stellar accretion rate, the disk surface density simplifies to
\begin{align}
    \Sigma_{\rm g, vis} &=  4.9 \times 10^3 \left( \frac{M_\star}{M_\odot} \right)^{7/5} \left( \frac{r}{\rm 1 \, au} \right)^{-3/5} \nonumber \\ 
     &\quad \times \left( \frac{\alpha_{\rm t}}{3 \cdot 10^{-4}} \right)^{-4/5}  {\rm g \, cm^{-2}} \\
    \Sigma_{\rm g, irr} &=  8.3 \times 10^3  \left( \frac{M_\star}{M_\odot} \right)^{31/14} \left( \frac{r}{\rm 1 \, au} \right)^{-15/14} \nonumber \\
     &\quad \times \left( \frac{\alpha_{\rm t}}{3 \cdot 10^{-4}} \right)^{-1}  {\rm g \, cm^{-2}}.
    \label{eq:sigma_g_numeric}
\end{align}
Equipped with this disk model, we can proceed and calculate how the terminal planet masses in the runaway accretion regime depend on stellar mass and orbital distance.

\subsection{Giant planet mass dependence on stellar mass and orbital separation}

We aim to determine how giant planet masses in the runaway regime depend on stellar mass and orbital separation. Given the complex interplay between disk properties, planetary growth rates, and gap opening, our goal is to conduct both analytical and numerical calculations to shed light on the relationship between planet mass, stellar mass, and orbital separation. Gas accretion rates in the runaway regime depend on the thermal mass $\qth$ of a planet, which is obtained by equating the Bondi radius to the local disk scale height:
\begin{align}
     \qth &= \left( \frac{M_{\rm p}}{M_\star} \right)  \left( \frac{H_{\rm p}}{r_{\rm p}} \right)^{-3}, \nonumber \\
     q_{\rm th, vis} &= \left( \frac{M_{\rm p}}{11 \, M_\oplus} \right)  \left( \frac{M_\star}{M_\odot} \right)^{-23/20} \nonumber \\ 
     &\quad \times  \left( \frac{r_{\rm p}}{\rm 1 \, au} \right)^{-3/20}  \left( \frac{\alpha_{\rm t}}{3 \cdot 10^{-4}} \right)^{3/10}, \\
     q_{\rm th, irr} &= \left( \frac{M_{\rm p}}{5 \, M_\oplus} \right)  \left( \frac{M_\star}{M_\odot} \right)^{1/14} \left( \frac{r_{\rm p}}{\rm 1 \, au} \right)^{-6/7}.
      \label{eq:qth}
\end{align}
$M_{\rm p}$ is the planet's mass and the subscript $_{\rm p}$ marks the relevant quantities at the location of the planet. In the viscous regime, the thermal mass is strongly dependent on stellar mass while its dependence on orbital separation is weak. In the passively irradiated regime, the strengths of the dependencies flip. To calculate the growth of giant planets in the runaway accretion regime and determine the dependence of a giant planet's mass on its host star's mass, we use the accretion rates from \cite{Choksi2023}.
\begin{numcases}{\frac{\dot{M}_{\mathrm{p,in}}}{\rho_{\mathrm{g}}\,\Omega_{\mathrm{p}}\,r_{\rm p}^{3}} \simeq}
  C_1\, \qth^{2} \left(\frac{H_{\mathrm{p}}}{r_{\rm p}}\right)^{3}
    & $\qth \lesssim 0.3$ \label{eq:mdot_low} \\
  C_2\, \qth \left(\frac{H_{\mathrm{p}}}{r_{\rm p}}\right)^{3}
    & $0.3 \lesssim \qth \lesssim 10$ \label{eq:mdot_mid} \\
  C_3\, \qth^{2/3} \left(\frac{H_{\mathrm{p}}}{r_{\rm p}}\right)^{3}
    & $\qth \gtrsim 10$. \label{eq:mdot_high}
\end{numcases}
We adopt values of constants $(C_1, C_2, C_3) = (3.5, 4/3, 9/3^{2/3})$ that correspond to sink-free simulations from \cite{Choksi2023}. For a vertically isothermal disk, the gas density at the planet's location $\rho_{\rm g} = \Sigma_{\rm g} / (\sqrt{2 \pi} H_{\rm p})$. As a planet grows in mass, it opens a gap in the disk and the accretion rate scales with the density of the gas in the gap $\Sigma_{\mathrm{gap}}$, which is given by \citep{Kanagawa2016}
\begin{equation}
    \Sigma_{\mathrm{gap}} = \frac{\Sigma_{\rm g}}{1 + K'},
\end{equation}
where $\Sigma_{\rm g}$ is the unperturbed gas density and the factor $K'$ is
\begin{align}
    K' &= 0.04 \left(\frac{M_{\mathrm{p}}}{M_\star}\right)^2 \left(\frac{H_{\mathrm{p}}}{r_{\rm p}}\right)^{-5} \frac{1}{\alpha_{\mathrm{t}}} \nonumber \\
      &= 0.04 \, \qth^2 \left(\frac{H_{\mathrm{p}}}{r_{\rm p}}\right) \frac{1}{\alpha_{\mathrm{t}}} \nonumber \\
      & \approx 7 \, \left(\frac{\qth}{1}\right)^2 \, \left(\frac{H_{\mathrm{p}} / r_{\rm p}}{0.05}\right) \left(\frac{\alpha_{\mathrm{t}}}{3 \cdot 10^{-4}}\right)^{-1}
      \label{eq:K}
\end{align}
Expressing $K'$ in terms of $\qth$ allows us to simplify gas accretion equations in different regimes. Rewriting Equations~\eqref{eq:mdot_low}--\eqref{eq:mdot_high} in terms of the thermal mass $\qth$ and including the effect of gap opening by the growing planet, we get
\begin{equation}
  \dot{q}_{\mathrm{th}} = \frac{G \rho_{\rm g}}{\Omega_{\mathrm{p}}} \, \frac{1}{1 + K'}\; C_i\,\qth^{a_i},
  \label{eq:qdot}
\end{equation}
where the subscript $i$ represents the three $\qth$ regimes for planetary growth rates. The term $G \rho_{\rm g}$ can also be rewritten as $1/t_{\rm ff}^2$, i.e., the local free fall time of the gas, ignoring an order unity constant. Equation~\ref{eq:qdot} assumes that stellar mass and the disk aspect ratio do not evolve with time. We solve this piecewise in each of these three regimes below. 

To obtain analytical solutions, we simplify the $1 / (1 + K')$ term in the different regimes\footnote{Equation~\ref{eq:qdot} can be solved analytically without simplifying the $1 / (1 + K')$ term. However, the solutions provide implicit expressions for $\qth$ that cannot always be converted into explicit functions of time.}. For $\qth < 0.3$, $K' \lesssim 0.7 $ for $\alpha_{\mathrm{t}} = 3 \cdot 10^{-4}$ and $H_{\mathrm{p}}/r_{\rm p} = 0.05$ and therefore to a good approximation, $K'$ in the denominator can be ignored in this regime. Stronger turbulence would improve this approximation. Disk aspect ratio is not expected to exceed $\sim 0.1$ so the effect of higher values of $H_{\mathrm{p}}/r_{\rm p}$ is capped at a factor of $\sim 2$. In the third regime when $\qth \gtrsim 10$, $K' \gtrsim 7 \times 10^2$ for fiducial values of $\alpha_{\mathrm{t}}$ and $H_{\mathrm{p}}/r_{\rm p}$ and therefore to a very good approximation, $1 / (1 + K') = 1/K'$. Even at higher turbulence, $K' \gg 1$. In the intermediate regime ($0.3 \lesssim \qth \lesssim 10$), fiducially $0.7 \lesssim K' \lesssim 7 \times 10^2$. Given that for most values of $\qth$ in the range $0.3 - 10 $, $K' > 1$, we elect to simplify $1 / (1 + K')$ as $ 1/K'$. 

\begin{figure*}
    \centering
    \includegraphics[width=0.49\linewidth]{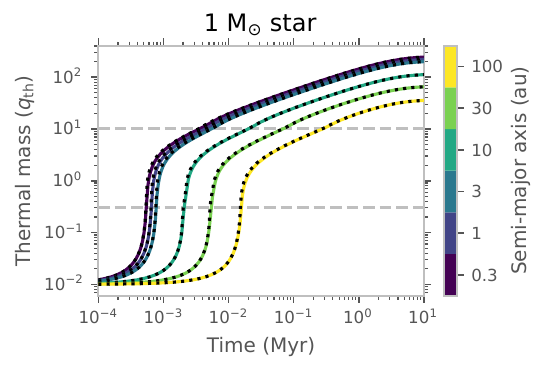}
    \includegraphics[width=0.49\linewidth]{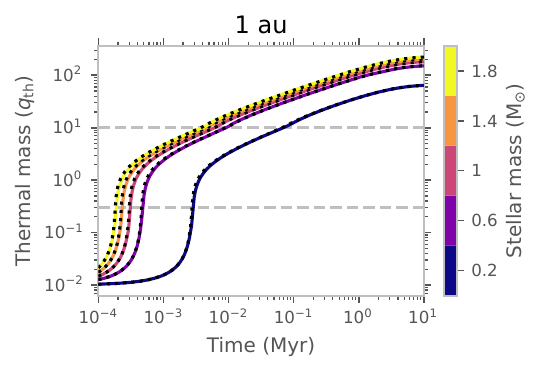}
    \caption{Thermal mass $\qth$ as a function of time for a range of semi-major axes at a fixed stellar mass (left) and for a range of stellar masses at a fixed semi-major axis (right). Solid colored lines and dotted black lines correspond to numerical and analytical solutions, respectively. The horizontal dashed gray lines mark $\qth = 0.3$ and $\qth = 10$, which separate different accretion regimes. The initial $\qth = 0.01$ is chosen for illustrative purposes only and does not reflect the mass at which runaway is triggered.}
    \label{fig:qth_semimajor_stellar}
\end{figure*}

With these approximations, $\qth$ can be analytically calculated as a function of time using the following expressions:
\begin{numcases}{\qth = }
  \frac{q_{\mathrm{th},0}}{1 - C_1 b_1 q_{\mathrm{th},0}\,(t - t_0)}
    & $\qth \lesssim 0.3$ \label{eq:q_low} \\
  \sqrt{q_{\mathrm{th},1}^{2} + 2\, C_2 b_{23}\,(t - t_1)}
    & $0.3 \lesssim \qth \lesssim 10$ \label{eq:q_mid} \\
  \left[q_{\mathrm{th},2}^{7/3} + \frac{7}{3}\, C_3 b_{23}\,(t - t_2)\right]^{3/7}
    & $\qth \gtrsim 10$. \label{eq:q_high}
\end{numcases}
Here, $q_{\mathrm{th},0}$ is the initial thermal mass of the giant planet at time $t_0$, $q_{\mathrm{th},1} = 0.3$, $q_{\mathrm{th},2} = 10$, $t_1$ and $t_2$ are times at which $\qth = q_{\mathrm{th},1}$ and $\qth = q_{\mathrm{th},2}$, respectively. Regime 1 can be ignored if the initial thermal mass $q_{\mathrm{th},0} > q_{\mathrm{th},1}$, i.e., for initial planet mass $\gtrsim1. 5 \, M_\oplus (M_\star/M_\odot)^{-1/14} (r/1\,{\rm au})^{6/7}$. The coefficients $b_1 = G \rho_{\rm g} / \Omega_{\rm p}$ and $b_{23} = b_1 \alpha_{\rm t}  (0.04 \, H_{\rm p} / r_{\rm p})^{-1}$ 
characterize the rate at which $\qth$ increases. The rate coefficient $b_{23}$ will play an important role in setting the growth timescale and final planet masses in the second and third regimes and it is given by:
\begin{align}
    b_{23, {\rm vis}} &\approx 0.01 ~ {\rm year^{-1}} ~ \left( \frac{M_\star}{M_\odot} \right)^{4/5} \left( \frac{r}{\rm 1 \, au} \right)^{-1/5} \left( \frac{\alpha_{\rm t}}{3 \cdot 10^{-4}} \right)^{2/5} \label{eq:b23_vis} \\
    b_{23, {\rm irr}} &\approx 0.03 ~ {\rm year^{-1}} ~ \left( \frac{M_\star}{M_\odot} \right)^{17/7} \left( \frac{r}{\rm 1 \, au} \right)^{-8/7}. \label{eq:b23_irr}
\end{align}

We analyze the analytical solutions, which make a number of things apparent. First, $\qth$ rises rapidly when it is $< 0.3$. The time $t_1$ at which $\qth = q_{\rm th, 1} =  0.3$, assuming $t_0 = 0$, is
\begin{align}
    t_1 &= \frac{1}{b_1 C_1 q_{\rm th, 1}} \left( \frac{q_{\rm th, 1}}{q_{\rm th, 0}} - 1 \right) \\
    t_{1, {\rm vis}} & = 22 ~ {\rm years} ~ \left( \frac{M_\star}{M_\odot} \right)^{-17/20} \left( \frac{r}{\rm 1 \, au} \right)^{3/20} \nonumber \\
    & \quad \times \left( \frac{\alpha_{\rm t}} {3 \cdot 10^{-4}} \right)^{7/10} \left( \frac{0.3}{q_{\rm th, 0}} - 1 \right), \nonumber \\
    t_{1, {\rm irr}} & = 10 ~ {\rm years} ~\left( \frac{M_\star}{M_\odot} \right)^{-29/14} \left( \frac{r}{\rm 1 \, au} \right)^{6/7} \nonumber \\
    & \quad \times \left( \frac{\alpha_{\rm t}} {3 \cdot 10^{-4}} \right) \left( \frac{0.3}{q_{\rm th, 0}} - 1 \right). \nonumber
\end{align}
The time taken to reach $q_{\rm th, 1} = 0.3$ is so short that for all intents and purposes, any reasonable choice (planet mass needed to trigger runaway accretion) of initial $q_{\rm th, 0} < q_{\rm th, 1}$ does not leave any imprint on the final planet mass. Indeed, for a wide range of stellar masses and semi-major axes, we might actually expect runaway accretion to be triggered only for $\qth \gtrsim 0.3$.

In the second regime, growth tapers off as $t^{1/2}$. The time it takes for planets to grow from $q_{\rm th, 1} = 0.3$ to $q_{\rm th, 2} = 10$ can be calculated using Equation~\ref{eq:q_mid}:
\begin{align}
    (t_2 - t_1)_{\rm vis} &= 3.8 \times 10^3 ~ {\rm years} ~ \left( \frac{M_\star}{M_\odot} \right)^{-4/5} \left( \frac{r}{\rm 1 \, au} \right)^{1/5} \nonumber \\
    &\quad \times \left( \frac{\alpha_{\rm t}} {3 \cdot 10^{-4}} \right)^{-2/5}  \\
    (t_2 - t_1)_{\rm irr} &= 1.3 \times 10^3~ {\rm years} ~ \left( \frac{M_\star}{M_\odot} \right)^{-17/7} \left( \frac{r}{\rm 1 \, au} \right)^{8/7}.
    \label{eq:phase_2}
\end{align}
This is an uncomfortably short duration, especially at $1-10$ au for stars more massive than M stars, because we know that many observed giant planets likely have $\qth \sim 0.3 - 10$. Taken at face value, such rapid growth of $\qth $ from 0.3 to 10 and the preponderance of planets in this $\qth$ range would require fine-tuning of the time available for runaway accretion. The actual accretion rate must be significantly slower than the maximum accretion rate from \cite{Choksi2023} used in this work. At the lowest stellar masses and at large orbital distances, this estimate of $t_2 - t_1$ starts approaching the disk lifetime and would need to be modified to account for disk dissipation (see below).

In the third regime, planetary growth tapers off as $t^{3/7}$, nearly at the same rate as the second regime. In the absence of disk dispersal (as assumed in these solutions), giant planets would continue growing indefinitely even though they open deeper gaps as time goes on. Terminating giant planet growth therefore requires the dispersal of protoplanetary disks. For exponential decay of disk density, one can simply replace $(t - t_i)$ in Equations~\ref{eq:q_low} -- \ref{eq:q_high} with $t_{\rm d} (e^{-t_i / t_{\rm d}} - e^{-t/ t_{\rm d}})$, where $t_{\rm d}$ is the dissipation timescale. For numerical estimates, we adopt $t_{\rm d} = 3$ Myr and a total disk lifetime $t_{\rm end} = 10$ Myr. Figure~\ref{fig:qth_semimajor_stellar} shows a comparison of the numerical solutions to Equations~\ref{eq:mdot_low} -- \ref{eq:mdot_high} and the analytical solutions obtained using Equations~\ref{eq:q_low} -- \ref{eq:q_high} (with $t - t_i$ replaced by $t_{\rm d} (e^{-t_i / t_{\rm d}} - e^{-t/ t_{\rm d}})$). The agreement is excellent and the two solutions differ noticeably only for values of $\qth$ for which $K' \sim 1$, as expected. The weaker dependencies of $t_1$ and $t_2 - t_1$ on $r$ and $M_\star$ in the viscously heated regions are manifest in Figure~\ref{fig:qth_semimajor_stellar}, where the curves corresponding to this regime are much closer together in $\qth - t$ space compared to growth curves in the passively heated regions.

Given that the growth of $\qth$ is extremely rapid in the first regime, we expect planets to stall in their growth only in the second or third regimes. Stellar mass dependence of $\qth$ is encapsulated in the coefficient $b_{23}$ that determines the rate at which $\qth$ grows. Planetary growth stalls in the second regime if 
\begin{align}
    \frac{q_{\rm th, 2}^2 - q_{\rm th, 1}^2}{2 C_2 b_{23}} &> t_{\rm d} (e^{-t_1/t_{\rm d}} - e^{-t/t_{\rm d}}) \\
    \frac{q_{\rm th, 2}^2 - q_{\rm th, 1}^2}{2 C_2 b_{23}} &\gtrsim t_{\rm d},
\end{align}
where the second expression is true in the limit of $t_1 \ll t_{\rm d}$ and $t \rightarrow \infty$. For a given $t_{\rm d}$, this expression gives us a relation between $r$ and $M_\star$ for which growth stalls in the second regime
\begin{equation}
     r \gtrsim 6.6 ~ {\rm au} ~ \left( \frac{M_\star}{0.1~M_\odot} \right)^{17/8} \left( \frac{t_{\rm d}}{\rm 3~Myr} \right)^{7/8}.
    \label{eq:r_constraint_regime_2}
\end{equation}
Here, we have limited ourselves to the passively irradiated part of the disk (compare to $r_{\rm vis \rightarrow irr}$, Equation~\ref{eq:r_vis_irr}) where growth in the second regime can be slow enough to stall before $\qth$ reaches $q_{\rm th,2} = 10$. In this regime, the final thermal mass is approximately given by
\begin{align}
    q_{\rm th, max} & \approx \left(2 C_2 b_{23} t_{\rm d} \right)^{1/2} \\
    & \approx 7.9 \left( \frac{M_\star}{0.1~M_\odot} \right)^{17/14} \left( \frac{r}{\rm 10 \, au} \right)^{-4/7} \left( \frac{t_{\rm d}}{\rm 3 \, Myr} \right)^{1/2},
\end{align}
which corresponds to a planet mass of 
\begin{equation}
    M_{\rm p, max} \approx 1~M_{\rm Jup} \left( \frac{M_\star}{0.1~M_\odot} \right)^{8/7} \left( \frac{r}{\rm 10 \, au} \right)^{2/7} \left( \frac{t_{\rm d}}{\rm 3 \, Myr} \right)^{1/2}.
    \label{eq:Mp_max_regime_2}
\end{equation}

\begin{figure}
    \centering
    \includegraphics[width=\linewidth]{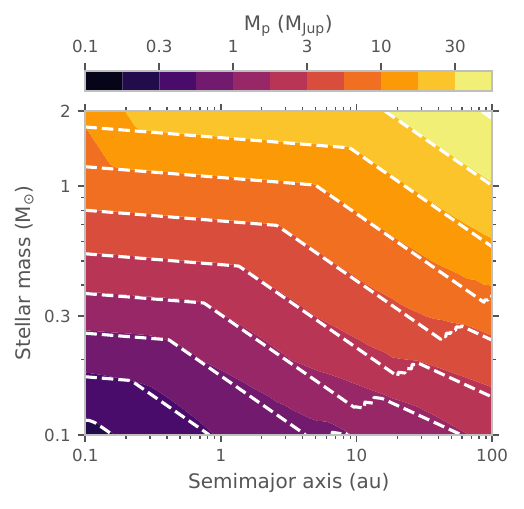}
    \caption{The maximum mass $M_{\rm p, max}$ expected from runaway accretion as a function of stellar mass and semi-major axis. Filled contours show numerical solutions. The dashed lines correspond to simple limits on $M_{\rm p, max}$ given in Equations~\ref{eq:Mp_max_regime_2}, \ref{eq:Mp_max_vis} and \ref{eq:Mp_max_irr}, which show reasonable agreement with numerical solutions, especially given their simplicity. The filled contours deviate in the top left (high $M_\star$, small $r$) because we cap disk temperature at 2000 K but this regime is not relevant for our study. We note that full analytical solutions (not shown) match the numerical solutions almost perfectly.}
    \label{fig:Mp_max_mstar_semimajor}
\end{figure}

In the third regime, the analytical estimate of the final thermal mass of a giant planet then takes the following form
\begin{align}
    q_{\rm th, max} &\approx \left( \frac{7}{3} C_3 b_{23} t_{\rm d} \right)^{3/7} \\
    q_{\rm th, max, vis} &\approx 220 \, \left( \frac{M_\star}{M_\odot} \right)^{12/35} \left( \frac{r}{\rm 1 \, au} \right)^{-3/35} \nonumber \\
    &\quad \times \left( \frac{\alpha_{\rm t}} {3 \cdot 10^{-4}} \right)^{6/35} \left( \frac{t_{\rm d}}{\rm 3 \, Myr} \right)^{3/7} \\
    q_{\rm th, max, irr} &\approx 350 \, \left( \frac{M_\star}{M_\odot} \right)^{51/49} \left( \frac{r}{\rm 1 \, au} \right)^{-24/49} \left( \frac{t_{\rm d}}{\rm 3 \, Myr} \right)^{3/7}.
\end{align}
Here, we have assumed that $t_2 \ll t_{\rm d}$ and $t \rightarrow \infty$. This limit on the thermal mass translates to a maximum planet mass of
\begin{align}
    M_{\rm p, max, vis} &\approx 7.7 \, M_{\rm Jup} \left( \frac{M_\star}{M_\odot} \right)^{209/140} \left( \frac{r}{\rm 1 \, au} \right)^{9/140} \nonumber \\
    &\quad \times \left( \frac{\alpha_{\rm t}} {3 \cdot 10^{-4}} \right)^{-9/70} \left( \frac{t_{\rm d}}{\rm 3 \, Myr} \right)^{3/7} \label{eq:Mp_max_vis} \\
    M_{\rm p, max, irr} &\approx 5.5 \, M_{\rm Jup} \left( \frac{M_\star}{M_\odot} \right)^{95/98} \left( \frac{r}{\rm 1 \, au} \right)^{18/49} \left( \frac{t_{\rm d}}{\rm 3 \, Myr} \right)^{3/7}.
    \label{eq:Mp_max_irr}
\end{align}
These estimates of the maximum planet mass are typically independent of the mass of the planet at which runaway is triggered: the rapidity of the first regime and early phase of the second regime erase this dependence. In general, both $t_1$ and $t_2$ depend on stellar mass as well but for $t$ much greater than these times, their effect on the stellar mass dependence of $\qth$ is negligible. Satisfactorily, these estimates are also independent of or very weakly dependent on the poorly constrained turbulence strength $\alpha_{\rm t}$. Finally, we note that the accretion rates from \cite{Choksi2023} are hard upper limits as they are calculated using an isothermal equation of state and correspond to total incoming mass into the planet's gravitational sphere of influence. The $M_{\rm p, max}$ we calculate here therefore also provides a strong upper limit on the terminal planet mass due to runaway accretion. To summarize, $M_{\rm p, max}$ is given by Equation~\ref{eq:Mp_max_vis} in viscously heated regions. In passively irradiated regions, $M_{\rm p, max}$ is given by Equation~\ref{eq:Mp_max_irr} in general and by Equation~\ref{eq:Mp_max_regime_2} if growth stalls in the second regime (for orbital distances and stellar masses delineated in Equation~\ref{eq:r_constraint_regime_2}).

Figure~\ref{fig:Mp_max_mstar_semimajor} shows the maximum planet mass calculated numerically and using Equations~\ref{eq:Mp_max_regime_2}, \ref{eq:Mp_max_vis}, and \ref{eq:Mp_max_irr}. The agreement between the two approaches is reasonable, with the largest deviations occurring when the disk dissipation timescale is comparable to the time $t_2$ taken to reach $q_{\rm th, 2} = 10$. In the viscous regime, $M_{\rm p, max, vis}$ is approximately $\propto M_\star^{3/2}$ and almost independent of orbital distance. In contrast, $M_{\rm p, max}$ is approximately $\propto M_\star^1$ for the second and third accretion regimes in passively heated regions and rises with orbital distance. The linear dependence on $M_\star$ is a result of the strong dependence of the disk surface density on $M_\star$, which stems from the observed correlation $\dot{M}_\star \propto M_\star^2$ \citep{hartmannAccretionPreMainSequenceStars2016}. These dependencies of maximum giant planet mass, especially on stellar mass, will have a profound impact on the expected yield of Gaia. 

\subsection{Location of Giant Planet Occurrence Rate Peak}

Here, we consider processes that likely determine where giant planet occurrence peaks and whether we should expect the peak location to depend on stellar mass. Given the complexity of this problem and the uncertainties arising from our limited knowledge of factors that control giant planet core formation as well as planet migration, we limit ourselves to understanding how the occurrence rate peak might trend with stellar mass rather than attempting to determine the exact location of the peak based on different pieces of physics. To this end, we consider two competing propositions regarding the conditions needed to form a giant planet: i) formation of a massive core that can undergo runaway accretion \citep{Mizuno1980, Stevenson1982} and ii) the advantage of being beyond the water snowline, where gas accretion might be more rapid \citep{Venturini2015, Chachan2021} and massive core formation might be more likely \citep{Stevenson1988}. The location where massive cores form and where water condenses have different dependence on stellar mass, as we show below. This leads to differing predictions for the expected giant planet yield from Gaia.

Giant planets require the formation of massive cores that can accrete a gaseous envelope quickly enough to trigger runaway gas accretion while the protoplanetary disk is still extant. These cores may form either via pebble or planetesimal accretion. However, we do not consider planetesimal accretion for two reasons. First, the timescale to form giant planet cores by planetesimals can be prohibitively long, especially around lower mass stars for which the dynamical clock at a given distance is slower \citep{Laughlin_2004}. Second, we found that the orbital distance at which a core of fixed mass forms increases with decreasing stellar mass, further exacerbating the timescale problem. We consider where $\sim 15 \, M_\oplus$ cores might form via pebble accretion as a function of stellar mass. Our choice of core mass is inspired by the empirical mass-metallicity relation of transiting planets from \cite{Chachan2025b}. Planetary cores growing via pebble accretion eventually reach a mass threshold where they start significantly perturbing the gas disk around them, thereby altering the dynamics of pebbles. Early studies assumed that pebble accretion stops at this threshold (termed `pebble isolation mass') by trapping the pebbles in a pressure maximum \citep{Morbidelli2012, Bitsch2018}. However, more recent work points to pebbles leaking from this trap, with a large fraction likely going to the inner disk \citep{Drazkowska2019, Stammler2023, VanClepper2025, Huang2025}. Nonetheless, the pebble isolation mass remains a useful metric as the accretion of pebbles is certainly altered beyond this threshold, even if it does not cease completely. This threshold is set by a comparison of the growing core's Hill radius and the local disk scale height and as a result, it can be expressed in terms of the thermal mass $\qth$ \citep{Bitsch2018}:
\begin{align}
    q_{\text{pebble-iso}} & \approx 0.5 \, \qth \nonumber \\
    M_{\text{pebble-iso}, irr} & = 2.5 M_{\oplus} \left(\frac{M_{\star}}{M_{\odot}}\right)^{-1/14} \bigg(\frac{r}{\rm 1 \, au}\bigg)^{6/7},
    \label{eq:pebble_iso}
\end{align}
where we have dropped additional terms that depend on the disk's pressure gradient and turbulence strength as we are mostly interested in the stellar mass dependence of $M_{\text{pebble-iso}}$. The explicit expression presented above is also only valid in the passively irradiated regions as the $15~M_\oplus$ threshold is reached only in these regions for almost the entire range of stellar mass considered here. The location where $M_{\text{pebble-iso}} = 15 \, M_\oplus$ is
\begin{equation}
    r (M_{\text{pebble-iso}} = 15 \, M_\oplus) = 8~{\rm au} \, \left(\frac{M_{\star}}{M_{\odot}}\right)^{1/12}
    \label{eq:r_pebble_iso}
\end{equation}
and it depends very weakly on stellar mass. This is because the explicit linear dependence of $M_{\text{pebble-iso}, irr}$ on $M_\star$ is nearly canceled out by the implicit dependence of $H_{\rm g}/r$ on $M_\star$. If a $15~M_\oplus$ core is all that is required to make a giant planet, we might expect the orbital separation at which the giant planet occurrence rate peaks to be nearly independent of stellar mass.  

\begin{figure*}
    \centering
    \includegraphics[width=\linewidth]{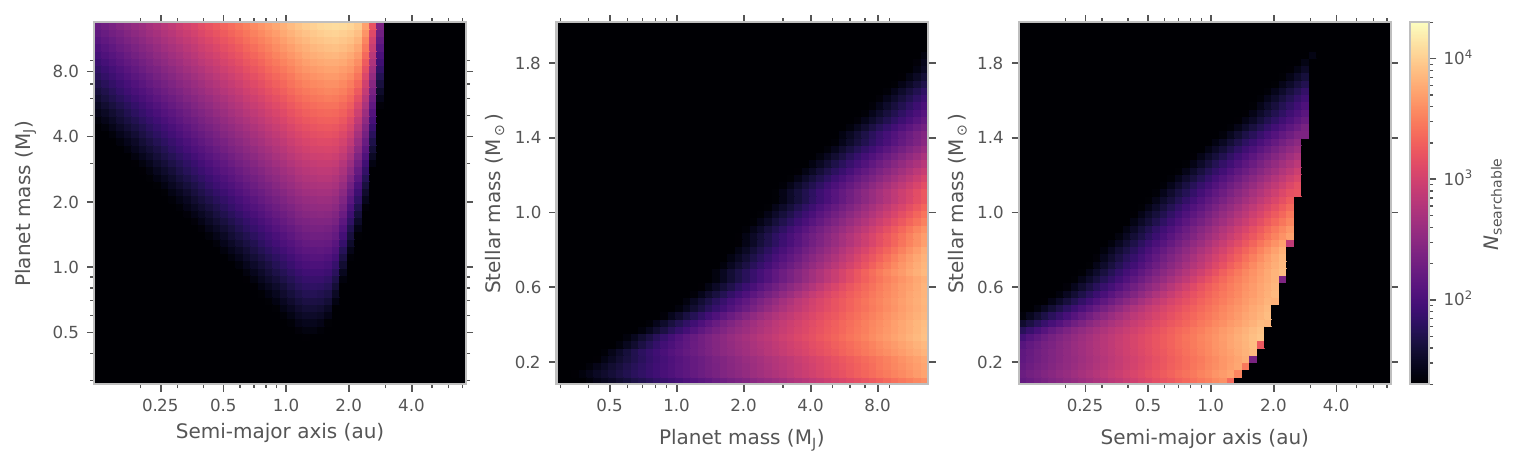}
    \caption{The number of searchable stars for Gaia DR4 ($\sim 1.1$ million) as a function of planet mass ($0.3 - 13~M_{\rm Jup}$), semi-major axis ($0.1 - 7$ au ), and stellar mass ($0.1 - 2~M_\odot$) from \cite{Lammers2025}. We multiply this data matrix with the planet occurrence rate to obtain the number of detectable planets (see Figure~\ref{fig:nplanets}).}
    \label{fig:nsearch}
\end{figure*}

The fate of cores that might turn into giant planets depends not only on how massive they are but also on the conditions in which these cores accrete nebular gas. Certain locations in the disk might also be conducive to massive core formation due to enhanced solid surface density. In this regard, the water snowline holds a special significance in the study of giant planet formation as it provides several favorable conditions for it. Although the condensation of water alone is not enough to increase the solid surface density enough to favor massive core formation (\citealt{Morbidelli2015}, if half of O is in water, condensation of water would add 20\% of the total metal mass available to the solids \citealt{Chachan2026}), the sublimation of water from dust that crosses the water snowline and its subsequent outward diffusion and recondensation onto dust could significantly boost the solid surface density and enable the formation of massive cores \citep{Stevenson1988, Ciesla2006, Schoonenberg2017, Drazkowska2017}. Alternatively, sublimation of water ice accreted via solids during gas accretion might enrich the planet's envelope, increase its mean molecular weight \citep{Stevenson1982, Hori2011}, cool down the envelope via latent heat release \citep{Venturini2015}, and trigger runaway accretion. In addition, radial drift of dust could leave the region beyond the water snowline dust-poor and accelerate gas accretion onto cores\footnote{The dustiness of the disk within the snowline depends on whether there is a significant difference in the fragmentation velocity of dust grains or turbulence strength across the snowline \citep{Chachan2021}.} \citep{Chachan2021}. Recent observational work presents tantalizing evidence that the location at which giant planet occurrence rate peaks is closer around M stars than around FGK stars \citep{Meyer2025}.

Assuming that water condenses at $\sim 170$ K in protoplanetary disk conditions and using Equations~\ref{eq:T_vis_simplified} and \ref{eq:T_irr_simplified}, we find that the location of the water snowline in viscously heated and passively heated regions is 
\begin{align}
    r_{\rm snowline, vis} &= 1.6  ~{\rm au} ~\bigg(\frac{M_\star}{M_{\odot}}\bigg)^{11/9} \left(\frac{\alpha_{\rm t}} {3 \cdot 10^{-4}} \right)^{-2/9}, \label{eq:r_snowline_vis} \\
    r_{\rm snowline, irr} &= 0.75 ~{\rm au} ~\bigg(\frac{M_\star}{M_{\odot}}\bigg)^{2/3},
    \label{eq:r_snowline_irr}
\end{align}
and the actual location for a given star and turbulence strength is given by $r_{\rm snowline}=$ max$(r_{\rm snowline, vis}, r_{\rm snowline, irr})$. For our fiducial value of $\alpha_{\rm t}$, $r_{\rm snowline} = r_{\rm vis \rightarrow irr}$ at $M_\star = 0.26~M_\odot$. Although the exact location of the snowline might be further out depending on the assumed thermal structure\footnote{This estimate ignores the weaker dependence of sublimation temperature on the disk pressure, which would slightly alter the stellar mass dependence of the snowline location.}, what is notable is the significantly stronger dependence of the snowline location on $M_\star$ compared to what we obtain for core formation by pebble accretion in Equation~\ref{eq:r_pebble_iso}. The water snowline shifts inwards as stellar mass decreases while the location at which a $15~M_\oplus$ core forms is nearly independent of stellar mass. This difference in the dependence of the locations of the water snowline and a constant $M_{\text{pebble-iso}}$ on $M_\star$ offers us a potent test of giant planet formation theories that we can observationally test with Gaia.

\section{Implications for Gaia}

Calculating the giant planet yield of the Gaia mission depends on the stellar and planetary properties that will yield measurable astrometric signals as well as the assumed occurrence rate of giant planets. \cite{Lammers2025} provided a semi-analytical estimate of the number of searchable stars as a function of stellar mass ($0.1 - 2~M_\odot$) and the mass ($0.3 - 13~M_{\rm Jup}$) and orbital separation ($0.1-7$ au) of giant planets. For Gaia DR4, they impose a limit of 4 years on the maximum period and adopt a detection threshold $N_\sigma = 1.5$ (per astrometric measurement). We use this estimate of searchable stars in our work (shown in Figure~\ref{fig:nsearch}), which cumulatively add up to $\sim 1.1$ million stars for Gaia DR4. Calculating the planet yield then simply involves multiplying this data cube of searchable stars with the planet occurrence rate. We will test the effect of our physically motivated changes to the planet occurrence rate on the planet yield of Gaia. 

Radial velocity surveys, in combination with constraints from direct imaging, provide the currently best available constraints on giant planet occurrence rate \citep{Fernandes2019, Nielsen2019, Wittenmyer2019, Fulton2021}. \cite{Lammers2025} provide a semi-analytical calculation for estimating Gaia's yield using the occurrence rate from \cite{Fulton2021}
\begin{equation}
    \frac{{\rm d}N_{\rm planets}}{{\rm d}N_\star \, \mathrm{d}\ln M_{\rm p} \, \mathrm{d}\ln r} = C \left( \frac{r}{1~{\rm au}} \right)^{\beta} (1 - e^{-(r/r_0)^\gamma}).
    \label{eq:Fulton_occurrence}
\end{equation}
The parameter $r_0$ corresponds to the location at which the occurrence rate peaks\footnote{\cite{Fulton2021} note that the occurrence rate jumps abruptly at 1 au (see also \citealt{Wittenmyer2019}) and stays roughly constant between $1-10$ au. This parametric model does not capture this behavior perfectly and likely underestimates the occurrence rate at 1 au by $\sim 10-20\%$.}. We adopt the reported median values for the fitted variables $C = 0.15$, $\beta = -0.86$, $r_0 = 3.6$ au, and $\gamma = 1.59$ since our goal is to show the effect of changing planet occurrence rate function rather than propagating uncertainties in these parameters. \cite{Lammers2025} add a linear $M_\star$ dependence to the constant $C$
\begin{equation}
    C(M_\star) = 0.15 \left( \frac{M_\star}{0.9~M_\odot} \right)
    \label{eq:LW2025_Mstar_dependence}
\end{equation}
to account for the lower overall occurrence rate of giant planets around lower mass stars. This is in reasonable agreement with observations \citep{Johnson2010, Clanton2014, Fulton2021, Pass2023} and theoretical expectations \citep[e.g.,][]{Chachan2023a}.

Notably, this form of the occurrence rate assumes that planet mass distribution does not depend either on stellar mass or orbital semi-major axis. In addition, the peak of the distribution is also independent of stellar mass. Massive super-Jupiters around K and M stars contribute significantly to the estimated Gaia yield in \cite{Lammers2025}. Our work suggests that these stars should harbor lower mass giants relative to more massive stars. To include the $M_\star$ and $r$ dependence of planet mass in the occurrence rate, we multiply Equation~\ref{eq:Fulton_occurrence} by 
\begin{equation}
    f(M_{\rm p}) = e^{-(M_{\rm p} / M_{\rm p, max})^2},
    \label{eq:f_Mpmax}
\end{equation}
where $M_{\rm p, max}$ is taken from Equations~\ref{eq:Mp_max_vis} and \ref{eq:Mp_max_irr}, as for the semi-major axes of interest, only the limiting mass in the third accretion regime is relevant. We choose this form for $f(M_{\rm p})$ to leave the occurrence rate for planets with $M_{\rm p} < M_{\rm p, max}$ relatively uniform while causing a rapid decline in the occurrence rate for planets more massive than $M_{\rm p, max}$. There is observational evidence to support a fall-off in planet occurrence rate at high planet masses: \cite{Fulton2021} find that the occurrence rate is flat in ln$M_{\rm p}$ space until $900~M_\oplus \sim 3~M_{\rm Jup}$ and falls off in the $\sim 3-10~M_{\rm Jup}$ mass range. Our values of $M_{\rm p, max}$, which are obtained from maximum accretion rates expected during runaway, sit squarely in this mass range.

\begin{figure*}
    \centering
    \includegraphics[width=\linewidth]{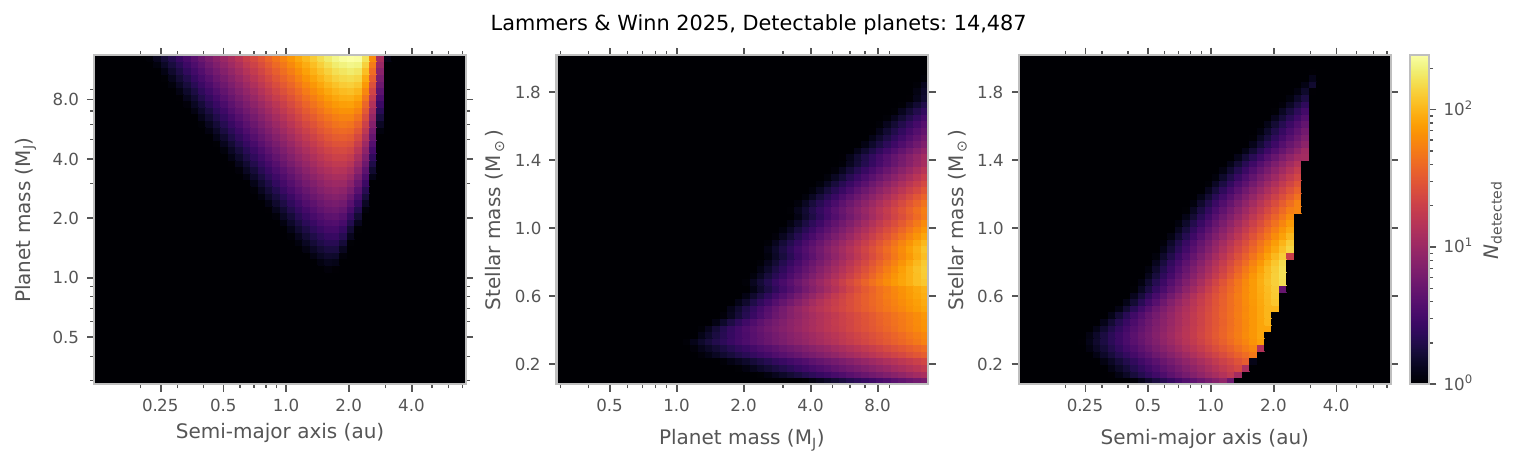}
    \includegraphics[width=\linewidth]{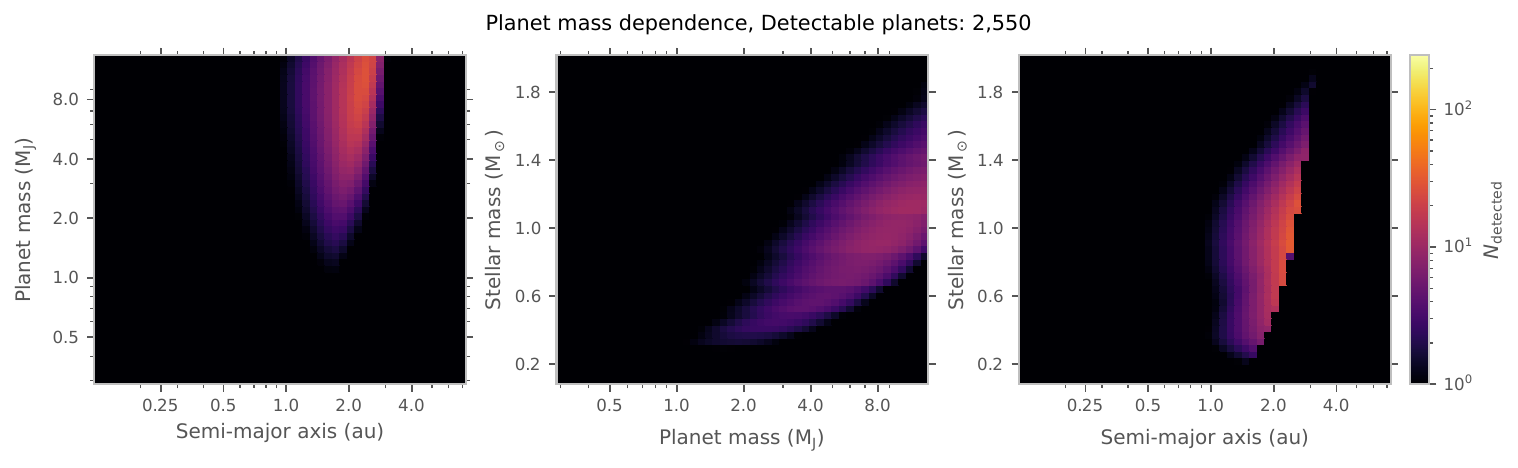}
    \includegraphics[width=\linewidth]{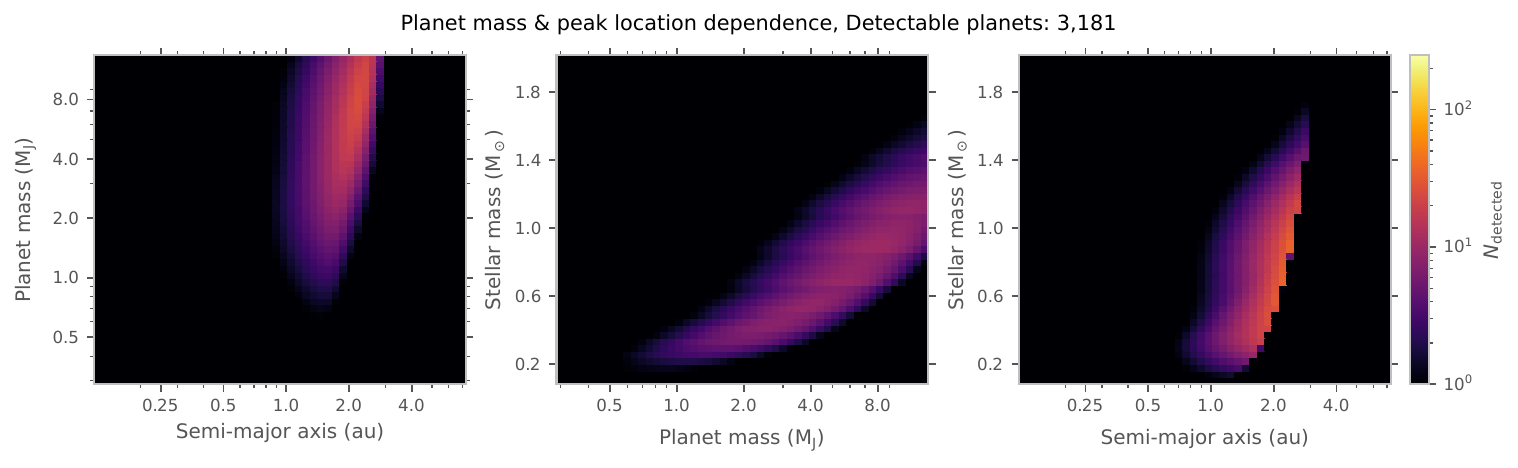}
    \caption{The number of detectable planets as a function of planet mass, semi-major axis, and stellar mass for three different planet occurrence rate functions. In the top panel, we use the planet occurrence rate from \cite{Fulton2021} and \cite{Lammers2025} (Equations~\ref{eq:Fulton_occurrence} and \ref{eq:LW2025_Mstar_dependence}), which only depends on orbital separation and is independent of planet mass. 
    In the middle panel, we include our limit on the maximum planet mass attainable via runaway accretion in the planet occurrence rate function (Equation~\ref{eq:f_Mpmax}), with $M_{\rm p, max} \approx 7.7~M_{\rm Jup} (M_\star/M_\odot)^{209/140} (r/{\rm 1~au})^{9/140}$ in viscously heated regions (Equation~\ref{eq:Mp_max_vis}) and $M_{\rm p, max} \approx 5.5~M_{\rm Jup} (M_\star/M_\odot)^{95/98} (r/{\rm 1~au})^{18/49}$ in passively irradiated regions (Equation~\ref{eq:Mp_max_irr}). These dependencies are driven by disk surface density and thermal structure and the dependence of stellar accretion rate on stellar mass. In the bottom panel, we vary the location of the occurrence rate peak $r_0 = 3.6~{\rm au} (M_\star / M_\odot)$ (Equation~\ref{eq:r_0_Mstar_dependence}) such that it trends similarly to the water snowline location with stellar mass. This corresponds to the scenario where the occurrence rate peak is set by physics related to the water snowline (e.g., role in core formation or accelerated gas accretion).}
    \label{fig:nplanets}
\end{figure*}

In addition, we compare Gaia's planet yield with and without an $M_\star$ dependence for the peak location $r_0$. Given that an $M_\star$-independent $r_0$ approximates the weak $M_\star^{1/12}$ dependence of the location at which a $15~M_\oplus$ core forms via pebble accretion sufficiently well, we only consider the $M_\star$ dependence expected from the location of the water snowline. Since $r_{\rm snowline} \propto M_\star^{11/9}$ in the viscous regime and $\propto M_\star^{2/3}$ in the passively heated regime (Equations~\ref{eq:r_snowline_vis} and \ref{eq:r_snowline_irr}), we choose to use an intermediate dependence of $M_\star^1$ for our calculations:
\begin{equation}
    r_0 (M_\star) = 3.6 ~{\rm au} ~\left( \frac{M_\star}{M_\odot} \right).
    \label{eq:r_0_Mstar_dependence}
\end{equation}
We retain the empirical constraint of $r_0 = 3.6$ au from \cite{Fulton2021}.

For any modification we make to the planet occurrence rate function, we renormalize the constant $C$ to match $N_{\rm planets}$ per 100 stars from the California Legacy Survey (CLS) for the default median values for the fitted parameters. In particular, we match the observed occurrence rate of $0.3 - 13~M_{\rm Jup}$ mass planets with semi-major axes $r = 2-8$ au over the CLS stellar mass distribution, which yields $15$ planets per 100 stars. This ensures that the difference in planet yield we report is purely due to the physically motivated modifications made to the occurrence rate.

\begin{figure*}
    \centering
    \includegraphics[width=0.32\linewidth]{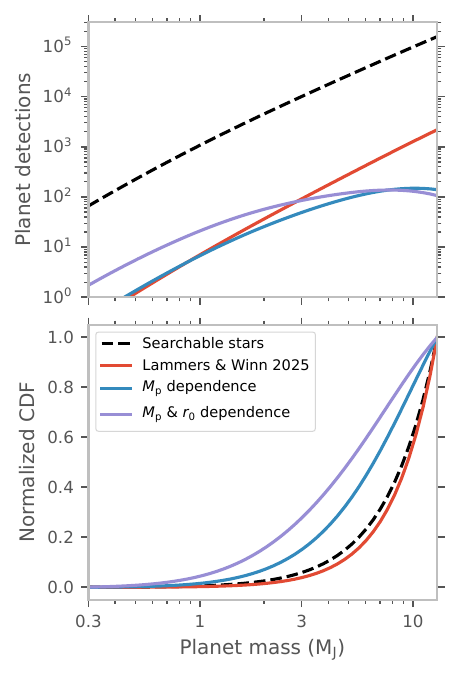}
    \includegraphics[width=0.32\linewidth]{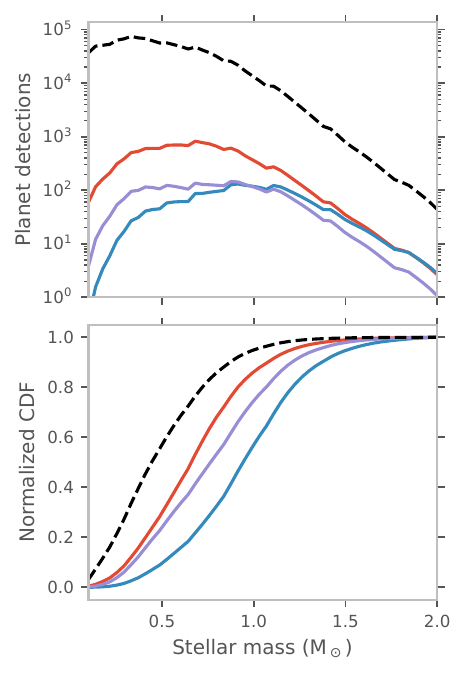}
    \includegraphics[width=0.32\linewidth]{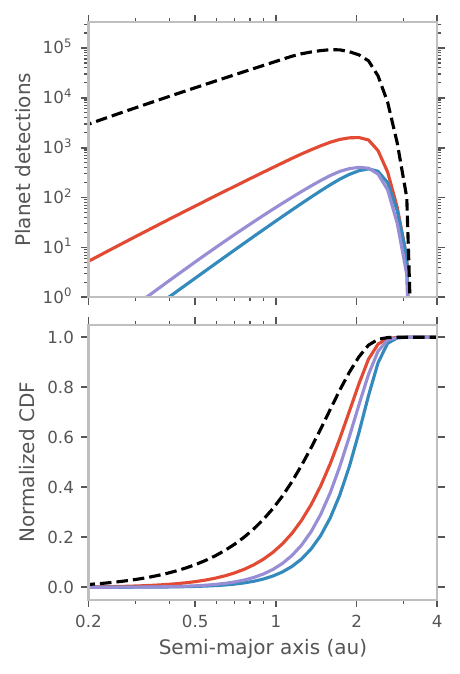}
    \caption{The number of planet detections (top) and the normalized cumulative distribution function for planet detections (bottom) as a function of planet mass (left), stellar mass (middle), and semi-major axis (right). Dashed black lines show the number of searchable stars and their CDF as a function of planet and host star properties. 
    The occurrence rate function (red) from \cite{Fulton2021} and \cite{Lammers2025} assumes that the planet mass distribution is independent of stellar mass and semi-major axis. Most (90\%) of the detected planets have masses $> 4.6 M_{\rm Jup}$ and half the detected planets orbit stars with mass $< 0.66~M_\odot$ in their model. Putting physically motivated $M_\star$ and $r$ dependent constraints obtained from runaway accretion on the planet mass has a dramatic effect on the expected planet detections (blue). In particular, the massive super-Jupiters around K and M stars that dominate the planet yield in \cite{Lammers2025} are no longer present. If the location of the occurrence rate peak $r_0 \propto M_\star$ (purple), the planet yield around K and M stars would make a partial recovery but the detected planets will have lower masses ($<6~M_{\rm Jup}$) than those in \cite{Lammers2025}.}
    \label{fig:detections_1d_plot}
\end{figure*}

Figure~\ref{fig:nplanets} shows the giant planets we expect to detect with Gaia by adopting the planet occurrence rate function from \cite{Fulton2021} and \cite{Lammers2025} (top panel), modifying it using Equation~\ref{eq:f_Mpmax} to account for the stellar mass and orbital separation dependence of maximum planet mass (middle panel), and adding a stellar mass dependence to the location of the occurrence rate peak (Equation~\ref{eq:r_0_Mstar_dependence}, bottom panel). Our physically motivated occurrence rate has a dramatic impact on the total predicted yield of giant planets in Gaia DR4. It reduces the expected number of detected planets from $\sim 14,500$ to $2,500 - 3,200$, depending on whether the peak of the occurrence rate is dependent on stellar mass or not. The largest difference is due to an absence of massive super-Jupiters ($> 3~M_{\rm Jup}$) orbiting K ($0.6 - 0.86~M_\odot$) and M stars ($< 0.6~M_\odot$) in our model: these constitute 79\% of the planets that are detected in \cite{Lammers2025} but are missing in our simulated planet sample. More broadly, i) half the missing planets orbit M stars and ii) half the missing planets have mass $\gtrsim 10~M_{\rm Jup}$. Moreover, planets with semi-major axis $\lesssim 0.4$ au are absent from our sample of detected planets: these again correspond to massive super-Jupiters around M stars that can produce detectable astrometric signals even at short orbital separations by virtue of their mass. Our estimates do not include massive super-Jupiters that may form by gravitational instability instead of runaway accretion. These planets could boost the number of massive super-Jupiters that are detected and modify the observed planet mass distribution by adding another peak in the distribution at $\gtrsim10~M_{\rm Jup}$ around lower mass stars. 

Figure~\ref{fig:detections_1d_plot} shows the 1D planet mass, host star mass, and semi-major axis distribution of the searchable stars and the expected planet detections for different planet occurrence rate functions. Comparing these distributions for searchable stars and planet detections allows us to see the imprint of the planet occurrence rate function. Since \cite{Lammers2025} assume that the occurrence rate is independent of planet mass, the planet mass CDF of searchable stars and detected planets lie almost on top of each other and the sample of detected planets is dominated by massive planets (90\% of the detected planets in \cite{Lammers2025} have mass $> 4.6~M_{\rm Jup}$). In contrast, the lower occurrence rate of giants at lower stellar masses and shorter orbital separations shifts the planet detection CDFs to higher stellar masses and larger orbital separations compared to CDFs of searchable stars. Placing a limit on planet mass as a function of stellar mass and orbital separation further sculpts the CDFs by shifting the distribution to lower planet masses, higher stellar masses, and larger orbital separations. Adding a stellar mass dependence to the location at which occurrence rate peaks partially offsets this shift in stellar mass and orbital separation by allowing more planets around K and M stars that are at closer separations to be detected. However, these additional planet detections shift the planet mass CDF further to the left as they typically correspond to planets that have lower masses ($<6~M_{\rm Jup}$). 

The planet occurrence rate functions suggested here have distinct effects on the 2D and 3D distribution of planets in the planet mass, semi-major axis, and stellar mass space. For example, the left boundary of the detected planet distribution in $M_{\rm p} - r$ space (left column in Figure~\ref{fig:nplanets}) changes sign such that $M_{\rm p}$ becomes positively correlated with $r$ at high planet masses. In the $M_\star - M_{\rm p}$ space (middle column, Figure~\ref{fig:nplanets}), the top boundary remains similar for the different occurrence rate models because it is set by the minimum astrometric displacement needed to detect a planet ($M_{\rm p} \propto M_\star$) but the bottom boundary is transformed by placing a limit on planet mass as a function of stellar mass and orbital separation. How far in $M_\star - M_{\rm p}$ space this `wedge' of detectable planets extends depends on whether the occurrence rate peak location depends on stellar mass or not. Similarly, in the $M_\star - r$ plane (right column, Figure~\ref{fig:nplanets}), the left sloped boundary is primarily set by the shortest orbital distance at which the most massive planet can be detected, i.e., a fixed astrometric displacement for the maximum planet mass allowed for a given stellar mass. For planet mass independent of stellar mass, this boundary would roughly follow $r \propto M_\star$. This effect still sets the left boundary for $M_\star \gtrsim 1~M_\odot$ when our planet mass and peak location dependence are added to the occurrence rate. However, for lower stellar masses, the boundary almost becomes a vertical cliff because lower mass stars host lower mass planets and such planets need to be at larger orbital separations to have the same astrometric displacement. Stellar mass dependence of the occurrence rate peak location can mold this cliff into a gentler $M_\star - r$ slope if the peak moves closer as $M_\star$ decreases. Overall, the 2D and 3D distribution of detected planets will provide insightful clues about processes that shape giant planet occurrence rate.

\section{Conclusions}

Our goal at the outset of the work was to provide physically motivated constraints on the giant planet occurrence rate. In particular, we wished to determine how planet mass distribution should depend on stellar mass and orbital separation and how the location at which giant planet occurrence rate peaks relates to different processes driving their formation. We focused on these dependencies because they have profound consequences for the planet yield expected from Gaia data releases.

In this work, we have provided analytical constraints on the maximum mass that giant planets should have as a function of their host star mass and orbital separation. This analysis revealed that planet mass is a strong function of stellar mass: in the final runaway accretion regime, $M_{\rm p, max}$ is approximately $\propto M_\star^{1.5} \, r^{0.06}$ and $M_\star^1 \, r^{0.37}$ in the viscously and passively heated regions, respectively. Terminating giant planet growth also requires the dispersal of the protoplanetary disk and final planet masses are $\propto t_{\rm d}^{3/7}$. Our constraint on $M_{\rm p, max}$ is commensurate with the planet mass range ($\sim 3-10~M_{\rm Jup}$) in which there is tentative evidence for a fall-off in giant planet occurrence rate \citep{Fulton2021}. We also highlighted how the stellar mass dependence of the location at which giant planet cores arise by pebble accretion and the location of the water snowline differ strongly from each other and therefore make distinct predictions about how the location of the giant planet occurrence rate peak should vary with stellar mass. 

Incorporating our physically motivated constraints into giant planet occurrence rate models dramatically alters the planet yield and leaves an imprint on the properties of the planets that we expect to find with Gaia. In particular, we do not expect massive super-Jupiters $\gtrsim 3~M_{\rm Jup}$ to exist or be common around K and M stars. As a result, our predicted yields for Gaia DR4 are a factor of $\sim 5$ smaller than the one reported in \cite{Lammers2025}. Importantly, the planet occurrence rate shapes the detected planet properties in planet mass, orbital separation, and stellar mass space. Therefore, the observed population can be used to infer the physical processes that drive terminal masses for giant planets and most favorable location for giant planet formation.

\begin{acknowledgments}
We thank Sarah Blunt for insightful conversations about Gaia DR4. J.J.F. acknowledges the support of a Simons Investigator grant.
\end{acknowledgments}

%




\bibliography{sample701}{}

@article{Choksi2023,
  title = {The maximum accretion rate of a protoplanet: how fast can runaway be?},
  volume = {525},
  copyright = {https://academic.oup.com/journals/pages/open\_access/funder\_policies/chorus/standard\_publication\_model},
  issn = {0035-8711, 1365-2966},
  shorttitle = {The maximum accretion rate of a protoplanet},
  url = {https://academic.oup.com/mnras/article/525/2/2806/7232547},
  doi = {10.1093/mnras/stad2269},
  language = {en},
  number = {2},
  urldate = {2026-08-04},
  journal = {Monthly Notices of the Royal Astronomical Society},
  author = {Choksi, Nick and Chiang, Eugene and Fung, Jeffrey and Zhu, Zhaohuan},
  month = aug,
  year = {2023},
  pages = {2806--2819},
}

@article{Kanagawa2016,
  title = {Mass constraint for a planet in a protoplanetary disk from the gap width},
  volume = {68},
  issn = {0004-6264},
  url = {https://doi.org/10.1093/pasj/psw037},
  doi = {10.1093/pasj/psw037},
  number = {3},
  urldate = {2026-10-27},
  journal = {Publications of the Astronomical Society of Japan},
  author = {Kanagawa, Kazuhiro D. and Muto, Takayuki and Tanaka, Hidekazu and Tanigawa, Takayuki and Takeuchi, Taku and Tsukagoshi, Takashi and Momose, Munetake},
  month = jun,
  year = {2016},
  pages = {43},
}

@article{Fulton2021,
  title = {California {Legacy} {Survey}. {II}. {Occurrence} of {Giant} {Planets} beyond the {Ice} {Line}},
  volume = {255},
  issn = {0067-0049},
  url = {https://ui.adsabs.harvard.edu/abs/2021ApJS..255...14F/abstract},
  doi = {10.3847/1538-4365/abfcc1},
  language = {en},
  number = {1},
  urldate = {2026-06-18},
  journal = {The Astrophysical Journal Supplement Series},
  author = {Fulton, Benjamin J. and Rosenthal, Lee J. and Hirsch, Lea A. and Isaacson, Howard and Howard, Andrew W. and Dedrick, Cayla M. and Sherstyuk, Ilya A. and Blunt, Sarah C. and Petigura, Erik A. and Knutson, Heather A. and Behmard, Aida and Chontos, Ashley and Crepp, Justin R. and Crossfield, Ian J. M. and Dalba, Paul A. and Fischer, Debra A. and Henry, Gregory W. and Kane, Stephen R. and Kosiarek, Molly and Marcy, Geoffrey W. and Rubenzahl, Ryan A. and Weiss, Lauren M. and Wright, Jason T.},
  month = jul,
  year = {2021},
  pages = {14},
}

@misc{Meyer2025,
  title = {Gas {Giant} and {Brown} {Dwarf} {Companions}: {Mass} {Ratio} and {Orbital} {Distributions} {From} {A} stars to {M} dwarfs},
  shorttitle = {Gas {Giant} and {Brown} {Dwarf} {Companions}},
  url = {https://scixplorer.org/abs/2025arXiv250805122M},
  doi = {10.48550/arXiv.2508.05122},
  urldate = {2026-11-21},
  publisher = {arXiv},
  author = {Meyer, Michael R. and Li, Yiting and Calissendorf, Per and Amara, Adam},
  month = aug,
  year = {2025},
  note = {ADS Bibcode: 2025arXiv250805122M},
}

@article{Chachan2023a,
  title = {Small {Planets} around {Cool} {Dwarfs}: {Enhanced} {Formation} {Efficiency} of {Super}-{Earths} around {M} {Dwarfs}},
  volume = {952},
  issn = {2041-8205, 2041-8213},
  shorttitle = {Small {Planets} around {Cool} {Dwarfs}},
  url = {https://iopscience.iop.org/article/10.3847/2041-8213/ace257},
  doi = {10.3847/2041-8213/ace257},
  language = {en},
  number = {1},
  urldate = {2026-03-13},
  journal = {The Astrophysical Journal Letters},
  author = {Chachan, Yayaati and Lee, Eve J.},
  month = jul,
  year = {2023},
  pages = {L20},
}

@article{Chiang1997,
  title = {Spectral {Energy} {Distributions} of {T} {Tauri} {Stars} with {Passive} {Circumstellar} {Disks}},
  volume = {490},
  issn = {0004-637X},
  url = {https://ui.adsabs.harvard.edu/abs/1997ApJ...490..368C/abstract},
  doi = {10.1086/304869},
  language = {en},
  number = {1},
  urldate = {2026-01-09},
  journal = {The Astrophysical Journal, Volume 490, Issue 1, pp. 368-376.},
  author = {Chiang, E. I. and Goldreich, P.},
  month = nov,
  year = {1997},
  pages = {368},
}

@article{Ida2016,
  author = {{Ida}, S. and {Guillot}, T. and {Morbidelli}, A.},
  title = {{The radial dependence of pebble accretion rates: A source of diversity in planetary systems. I. Analytical formulation}},
  journal = {\aap},
  year = {2016},
  month = jun,
  volume = {591},
  eid = {A72},
  pages = {A72},
  doi = {10.1051/0004-6361/201628099},
  archiveprefix = {arXiv},
  eprint = {1604.01291},
  primaryclass = {astro-ph.EP},
  adsurl = {https://ui.adsabs.harvard.edu/abs/2016A&A...591A..72I},
}

@article{hartmannAccretionPreMainSequenceStars2016,
  title = {Accretion onto {Pre}-{Main}-{Sequence} {Stars}},
  volume = {54},
  issn = {0066-4146, 1545-4282},
  url = {https://www.annualreviews.org/doi/10.1146/annurev-astro-081915-023347},
  doi = {10.1146/annurev-astro-081915-023347},
  language = {en},
  number = {1},
  urldate = {2024-10-21},
  journal = {Annual Review of Astronomy and Astrophysics},
  author = {Hartmann, Lee and Herczeg, Gregory and Calvet, Nuria},
  month = sep,
  year = {2016},
  pages = {135--180},
}

@article{Fernandes2019,
  author = {{Fernandes}, Rachel B. and {Mulders}, Gijs D. and {Pascucci}, Ilaria and {Mordasini}, Christoph and {Emsenhuber}, Alexandre},
  title = {{Hints for a Turnover at the Snow Line in the Giant Planet Occurrence Rate}},
  journal = {\apj},
  year = {2019},
  month = mar,
  volume = {874},
  number = {1},
  eid = {81},
  pages = {81},
  doi = {10.3847/1538-4357/ab0300},
  archiveprefix = {arXiv},
  eprint = {1812.05569},
  primaryclass = {astro-ph.SR},
  adsurl = {https://ui.adsabs.harvard.edu/abs/2019ApJ...874...81F},
}

@article{Laughlin_2004,
  title = {The Core Accretion Model Predicts Few Jovian-Mass Planets Orbiting Red Dwarfs},
  volume = {612},
  issn = {1538-4357},
  url = {http://dx.doi.org/10.1086/424384},
  doi = {10.1086/424384},
  number = {1},
  journal = {The Astrophysical Journal},
  publisher = {American Astronomical Society},
  author = {Laughlin, Gregory and Bodenheimer, Peter and Adams, Fred C.},
  year = {2004},
  month = Aug,
  pages = {L73–L76},
}

@article{Venturini2015,
  title = {Critical core mass for enriched envelopes: the role of {H}{\textless}{SUB}{\textgreater}2{\textless}/{SUB}{\textgreater}{O} condensation},
  volume = {576},
  issn = {0004-6361},
  shorttitle = {Critical core mass for enriched envelopes},
  url = {https://ui.adsabs.harvard.edu/abs/2015A&A...576A.114V/abstract},
  doi = {10.1051/0004-6361/201424008},
  language = {en},
  urldate = {2025-08-12},
  journal = {Astronomy and Astrophysics},
  author = {Venturini, J. and Alibert, Y. and Benz, W. and Ikoma, M.},
  month = apr,
  year = {2015},
  pages = {A114},
}

@article{Chachan2021,
  title = {Radial {Gradients} in {Dust}-to-gas {Ratio} {Lead} to {Preferred} {Region} for {Giant} {Planet} {Formation}},
  volume = {919},
  issn = {0004-637X, 1538-4357},
  url = {https://iopscience.iop.org/article/10.3847/1538-4357/ac0bb6},
  doi = {10.3847/1538-4357/ac0bb6},
  language = {en},
  number = {1},
  urldate = {2026-03-13},
  journal = {The Astrophysical Journal},
  author = {Chachan, Yayaati and Lee, Eve J. and Knutson, Heather A.},
  month = sep,
  year = {2021},
  pages = {63},
}

@article{Stevenson1988,
  author = {Stevenson, David J and Lunine, Jonathan I},
  journal = {Icarus},
  pages = {146--155},
  title = {Rapid Formation of Jupiter by Diffusive Redistribution of Water Vapor in the Solar Nebula},
  volume = {75},
  year = {1988},
}

@article{Mizuno1980,
  title = {Formation of the {Giant} {Planets}},
  volume = {64},
  issn = {0033-0684},
  url = {https://ui.adsabs.harvard.edu/abs/1980PThPh..64..544M/abstract},
  doi = {10.1143/PTP.64.544},
  language = {en},
  number = {2},
  urldate = {2025-05-23},
  journal = {Progress of Theoretical Physics},
  author = {Mizuno, H.},
  month = aug,
  year = {1980},
  pages = {544--557},
}

@article{Chachan2025b,
  title = {Revising the {Giant} {Planet} {Mass}–{Metallicity} {Relation}: {Deciphering} the {Formation} {Sequence} of {Giant} {Planets}},
  volume = {994},
  issn = {0004-637X, 1538-4357},
  shorttitle = {Revising the {Giant} {Planet} {Mass}–{Metallicity} {Relation}},
  url = {https://iopscience.iop.org/article/10.3847/1538-4357/ae0cbf},
  doi = {10.3847/1538-4357/ae0cbf},
  language = {en},
  number = {1},
  urldate = {2026-03-13},
  journal = {The Astrophysical Journal},
  author = {Chachan, Yayaati and Fortney, Jonathan J. and Ohno, Kazumasa and Thorngren, Daniel and Murray-Clay, Ruth},
  month = nov,
  year = {2025},
  pages = {43},
}

@article{Morbidelli2012,
  title = {Building {Terrestrial} {Planets}},
  volume = {40},
  issn = {0084-6597},
  url = {https://ui.adsabs.harvard.edu/abs/2012AREPS..40..251M/abstract},
  doi = {10.1146/annurev-earth-042711-105319},
  language = {en},
  number = {1},
  urldate = {2026-06-29},
  journal = {Annual Review of Earth and Planetary Sciences, vol. 40, issue 1, pp. 251-275},
  author = {Morbidelli, A. and Lunine, J. I. and O'Brien, D. P. and Raymond, S. N. and Walsh, K. J.},
  month = may,
  year = {2012},
  pages = {251},
}

@article{Bitsch2018,
  archiveprefix = {arXiv},
  arxivid = {1801.02341},
  author = {Bitsch, Bertram and Morbidelli, Alessandro and Johansen, Anders and Lega, Elena and Lambrechts, Michiel and Crida, Aur{\'{e}}lien},
  eprint = {1801.02341},
  journal = {eprint arXiv:1801.02341},
  month = {jan},
  title = {{Pebble isolation mass --- scaling law and implications for the formation of super-Earths and gas giants}},
  url = {http://arxiv.org/abs/1801.02341},
  year = {2018},
}

@article{Drazkowska2019,
  title = {Including {Dust} {Coagulation} in {Hydrodynamic} {Models} of {Protoplanetary} {Disks}: {Dust} {Evolution} in the {Vicinity} of a {Jupiter}-mass {Planet}},
  volume = {885},
  issn = {0004-637X},
  shorttitle = {Including {Dust} {Coagulation} in {Hydrodynamic} {Models} of {Protoplanetary} {Disks}},
  url = {https://ui.adsabs.harvard.edu/abs/2019ApJ...885...91D},
  doi = {10.3847/1538-4357/ab46b7},
  urldate = {2025-03-01},
  journal = {The Astrophysical Journal},
  publisher = {IOP},
  author = {Drazkowska, Joanna and Li, Shengtai and Birnstiel, Til and Stammler, Sebastian M. and Li, Hui},
  month = nov,
  year = {2019},
  note = {ADS Bibcode: 2019ApJ...885...91D},
  pages = {91},
}

@article{Stammler2023,
  title = {Leaky dust traps: {How} fragmentation impacts dust filtering by planets},
  volume = {670},
  issn = {0004-6361},
  shorttitle = {Leaky dust traps},
  url = {https://ui.adsabs.harvard.edu/abs/2023A&A...670L...5S},
  doi = {10.1051/0004-6361/202245512},
  urldate = {2025-03-01},
  journal = {Astronomy and Astrophysics},
  author = {Stammler, Sebastian Markus and Lichtenberg, Tim and Drazkowska, Joanna and Birnstiel, Tilman},
  month = feb,
  year = {2023},
  note = {ADS Bibcode: 2023A\&A...670L...5S},
  pages = {L5},
}

@article{VanClepper2025,
  title = {Three-dimensional {Transport} of {Solids} in a {Protoplanetary} {Disk} {Containing} a {Growing} {Giant} {Planet}},
  volume = {980},
  issn = {0004-637X},
  url = {https://ui.adsabs.harvard.edu/abs/2025ApJ...980..201V},
  doi = {10.3847/1538-4357/ada8a4},
  urldate = {2025-02-28},
  journal = {The Astrophysical Journal},
  publisher = {IOP},
  author = {Van Clepper, Eric and Price, Ellen M. and Ciesla, Fred J.},
  month = feb,
  year = {2025},
  note = {ADS Bibcode: 2025ApJ...980..201V},
  pages = {201},
}

@article{Huang2025,
  author = {{Huang}, Pinghui and {Yu}, Fangyuan and {Lee}, Eve J. and {Dong}, Ruobing and {Bai}, Xue-Ning},
  title = {{Leaky Dust Traps in Planet-embedded Protoplanetary Disks}},
  journal = {\apj},
  year = {2025},
  month = jul,
  volume = {988},
  number = {1},
  eid = {94},
  pages = {94},
  doi = {10.3847/1538-4357/addd1f},
  archiveprefix = {arXiv},
  eprint = {2503.19026},
  primaryclass = {astro-ph.EP},
  adsurl = {https://ui.adsabs.harvard.edu/abs/2025ApJ...988...94H},
}

@article{Stevenson1982,
  title = {Formation of the giant planets},
  volume = {30},
  issn = {0032-0633},
  url = {https://ui.adsabs.harvard.edu/abs/1982P%26SS...30..755S/abstract},
  doi = {10.1016/0032-0633(82)90108-8},
  language = {en},
  number = {8},
  urldate = {2025-05-23},
  journal = {Planetary and Space Science, Volume 30, Issue 8, p. 755-764.},
  author = {Stevenson, D. J.},
  month = aug,
  year = {1982},
  pages = {755},
}

@article{Chachan2026,
  title = {Quantitative {Links} between {Formation} and {Atmospheric} {Composition} for {Giant} {Planets}},
  volume = {1010},
  issn = {2041-8205, 2041-8213},
  url = {https://iopscience.iop.org/article/10.3847/2041-8213/aeaf59},
  doi = {10.3847/2041-8213/aeaf59},
  language = {en},
  number = {2},
  urldate = {2026-11-13},
  journal = {The Astrophysical Journal Letters},
  author = {Chachan, Yayaati and Murray-Clay, Ruth and Fortney, Jonathan J.},
  month = oct,
  year = {2026},
  pages = {L25},
}

@article{Hori2011,
  title = {Gas giant formation with small cores triggered by envelope pollution by icy planetesimals: {Envelope} pollution of gas giants by icy planetesimals},
  volume = {416},
  issn = {00358711},
  shorttitle = {Gas giant formation with small cores triggered by envelope pollution by icy planetesimals},
  url = {https://academic.oup.com/mnras/article-lookup/doi/10.1111/j.1365-2966.2011.19140.x},
  doi = {10.1111/j.1365-2966.2011.19140.x},
  language = {en},
  number = {2},
  urldate = {2025-08-12},
  journal = {Monthly Notices of the Royal Astronomical Society},
  author = {Hori, Y. and Ikoma, M.},
  month = sep,
  year = {2011},
  pages = {1419--1429},
}

@article{Morbidelli2015,
  title = {The great dichotomy of the {Solar} {System}: {Small} terrestrial embryos and massive giant planet cores},
  volume = {258},
  issn = {00191035},
  shorttitle = {The great dichotomy of the {Solar} {System}},
  url = {https://linkinghub.elsevier.com/retrieve/pii/S0019103515002456},
  doi = {10.1016/j.icarus.2015.06.003},
  language = {en},
  urldate = {2026-07-26},
  journal = {Icarus},
  author = {Morbidelli, A. and Lambrechts, M. and Jacobson, S. and Bitsch, B.},
  month = sep,
  year = {2015},
  pages = {418--429},
}

@article{Schoonenberg2017,
  title = {Planetesimal formation near the snowline: in or out?},
  volume = {602},
  issn = {1432-0746},
  url = {http://dx.doi.org/10.1051/0004-6361/201630013},
  doi = {10.1051/0004-6361/201630013},
  journal = {Astronomy \& Astrophysics},
  publisher = {EDP Sciences},
  author = {Schoonenberg, Djoeke and Ormel, Chris W.},
  year = {2017},
  month = May,
  pages = {A21},
}

@article{Lammers2025,
  title = {On the Exoplanet Yield of Gaia Astrometry},
  volume = {171},
  issn = {1538-3881},
  url = {http://dx.doi.org/10.3847/1538-3881/ae21de},
  doi = {10.3847/1538-3881/ae21de},
  number = {1},
  journal = {The Astronomical Journal},
  publisher = {American Astronomical Society},
  author = {Lammers, Caleb and Winn, Joshua N.},
  year = {2025},
  month = Dec,
  pages = {18},
}

@article{Nielsen2019,
  title = {The {Gemini} {Planet} {Imager} {Exoplanet} {Survey}: {Giant} {Planet} and {Brown} {Dwarf} {Demographics} from 10 to 100 au},
  volume = {158},
  issn = {0004-6256},
  shorttitle = {The {Gemini} {Planet} {Imager} {Exoplanet} {Survey}},
  url = {https://ui.adsabs.harvard.edu/abs/2019AJ....158...13N/abstract},
  doi = {10.3847/1538-3881/ab16e9},
  language = {en},
  number = {1},
  urldate = {2026-09-25},
  journal = {The Astronomical Journal},
  author = {Nielsen, Eric L. and De Rosa, Robert J. and Macintosh, Bruce and Wang, Jason J. and Ruffio, Jean-Baptiste and Chiang, Eugene and Marley, Mark S. and Saumon, Didier and Savransky, Dmitry and Ammons, S. Mark and Bailey, Vanessa P. and Barman, Travis and Blain, Célia and Bulger, Joanna and Burrows, Adam and Chilcote, Jeffrey and Cotten, Tara and Czekala, Ian and Doyon, Rene and Duchêne, Gaspard and Esposito, Thomas M. and Fabrycky, Daniel and Fitzgerald, Michael P. and Follette, Katherine B. and Fortney, Jonathan J. and Gerard, Benjamin L. and Goodsell, Stephen J. and Graham, James R. and Greenbaum, Alexandra Z. and Hibon, Pascale and Hinkley, Sasha and Hirsch, Lea A. and Hom, Justin and Hung, Li-Wei and Dawson, Rebekah Ilene and Ingraham, Patrick and Kalas, Paul and Konopacky, Quinn and Larkin, James E. and Lee, Eve J. and Lin, Jonathan W. and Maire, Jérôme and Marchis, Franck and Marois, Christian and Metchev, Stanimir and Millar-Blanchaer, Maxwell A. and Morzinski, Katie M. and Oppenheimer, Rebecca and Palmer, David and Patience, Jennifer and Perrin, Marshall and Poyneer, Lisa and Pueyo, Laurent and Rafikov, Roman R. and Rajan, Abhijith and Rameau, Julien and Rantakyrö, Fredrik T. and Ren, Bin and Schneider, Adam C. and Sivaramakrishnan, Anand and Song, Inseok and Soummer, Remi and Tallis, Melisa and Thomas, Sandrine and Ward-Duong, Kimberly and Wolff, Schuyler},
  month = jul,
  year = {2019},
  pages = {13},
}

@article{Johnson2010,
  title = {Giant Planet Occurrence in the Stellar Mass-Metallicity Plane},
  volume = {122},
  issn = {1538-3873},
  url = {http://dx.doi.org/10.1086/655775},
  doi = {10.1086/655775},
  number = {894},
  journal = {Publications of the Astronomical Society of the Pacific},
  publisher = {IOP Publishing},
  author = {Johnson, John Asher and Aller, Kimberly M. and Howard, Andrew W. and Crepp, Justin R.},
  year = {2010},
  month = Aug,
  pages = {905–915},
}

@article{Clanton2014,
  title = {SYNTHESIZING EXOPLANET DEMOGRAPHICS FROM RADIAL VELOCITY AND MICROLENSING SURVEYS. II. THE FREQUENCY OF PLANETS ORBITING M DWARFS},
  volume = {791},
  issn = {1538-4357},
  url = {http://dx.doi.org/10.1088/0004-637X/791/2/91},
  doi = {10.1088/0004-637x/791/2/91},
  number = {2},
  journal = {The Astrophysical Journal},
  publisher = {American Astronomical Society},
  author = {Clanton, Christian and Gaudi, B. Scott},
  year = {2014},
  month = July,
  pages = {91},
}

@article{Pass2023,
  title = {Mid-to-late M Dwarfs Lack Jupiter Analogs},
  volume = {166},
  issn = {1538-3881},
  url = {http://dx.doi.org/10.3847/1538-3881/acd349},
  doi = {10.3847/1538-3881/acd349},
  number = {1},
  journal = {The Astronomical Journal},
  publisher = {American Astronomical Society},
  author = {Pass, Emily K. and Winters, Jennifer G. and Charbonneau, David and Irwin, Jonathan M. and Latham, David W. and Berlind, Perry and Calkins, Michael L. and Esquerdo, Gilbert A. and Mink, Jessica},
  year = {2023},
  month = June,
  pages = {11},
}

@article{Chachan2022,
  title = {Kepler-167e as a {Probe} of the {Formation} {Histories} of {Cold} {Giants} with {Inner} {Super}-{Earths}},
  volume = {926},
  issn = {0004-637X, 1538-4357},
  url = {https://iopscience.iop.org/article/10.3847/1538-4357/ac3ed6},
  doi = {10.3847/1538-4357/ac3ed6},
  language = {en},
  number = {1},
  urldate = {2026-03-13},
  journal = {The Astrophysical Journal},
  author = {Chachan, Yayaati and Dalba, Paul A. and Knutson, Heather A. and Fulton, Benjamin J. and Thorngren, Daniel and Beichman, Charles and Ciardi, David R. and Howard, Andrew W. and Van Zandt, Judah},
  month = feb,
  year = {2022},
  pages = {62},
}

@article{Pollack1996a,
  title = {Formation of the {Giant} {Planets} by {Concurrent} {Accretion} of {Solids} and {Gas}},
  volume = {124},
  copyright = {https://www.elsevier.com/tdm/userlicense/1.0/},
  issn = {00191035},
  url = {https://linkinghub.elsevier.com/retrieve/pii/S0019103596901906},
  doi = {10.1006/icar.1996.0190},
  language = {en},
  number = {1},
  urldate = {2026-07-27},
  journal = {Icarus},
  author = {Pollack, James B. and Hubickyj, Olenka and Bodenheimer, Peter and Lissauer, Jack J. and Podolak, Morris and Greenzweig, Yuval},
  month = nov,
  year = {1996},
  pages = {62--85},
}

@article{Fischer2005,
  title = {The Planet‐Metallicity Correlation},
  volume = {622},
  issn = {1538-4357},
  url = {http://dx.doi.org/10.1086/428383},
  doi = {10.1086/428383},
  number = {2},
  journal = {The Astrophysical Journal},
  publisher = {American Astronomical Society},
  author = {Fischer, Debra A. and Valenti, Jeff},
  year = {2005},
  month = Apr,
  pages = {1102–1117},
}

@article{Ginzburg2019,
  title = {The end of runaway: how gap opening limits the final masses of gas giants},
  volume = {487},
  issn = {0035-8711},
  shorttitle = {The end of runaway},
  url = {https://doi.org/10.1093/mnras/stz1322},
  doi = {10.1093/mnras/stz1322},
  number = {1},
  urldate = {2026-11-21},
  journal = {Monthly Notices of the Royal Astronomical Society},
  author = {Ginzburg, Sivan and Chiang, Eugene},
  month = jul,
  year = {2019},
  pages = {681--690},
}

@article{Rosenthal2020,
  title = {How consumption and repulsion set planetary gap depths and the final masses of gas giants},
  volume = {498},
  copyright = {https://academic.oup.com/journals/pages/open\_access/funder\_policies/chorus/standard\_publication\_model},
  issn = {0035-8711, 1365-2966},
  url = {https://academic.oup.com/mnras/article/498/2/2054/5859965},
  doi = {10.1093/mnras/staa1721},
  language = {en},
  number = {2},
  urldate = {2026-11-21},
  journal = {Monthly Notices of the Royal Astronomical Society},
  author = {Rosenthal, M M and Chiang, E I and Ginzburg, S and Murray-Clay, R A},
  month = sep,
  year = {2020},
  pages = {2054--2067},
}

@article{Ciesla2006,
  title = {The evolution of the water distribution in a viscous protoplanetary disk},
  volume = {181},
  issn = {0019-1035},
  url = {https://scixplorer.org/abs/2006Icar..181..178C},
  doi = {10.1016/j.icarus.2005.11.009},
  urldate = {2026-11-29},
  journal = {Icarus},
  publisher = {Elsevier},
  author = {Ciesla, Fred J. and Cuzzi, Jeffrey N.},
  month = mar,
  year = {2006},
  note = {ADS Bibcode: 2006Icar..181..178C},
  pages = {178--204},
}

@article{Drazkowska2017,
  author = {Drazkowska, J. and Alibert, Y.},
  title = {Planetesimal formation starts at the snow line},
  journal = {Astronomy \& Astrophysics},
  year = {2017},
  volume = {608},
  issn = {1432-0746},
  url = {http://dx.doi.org/10.1051/0004-6361/201731491},
  doi = {10.1051/0004-6361/201731491},
  publisher = {EDP Sciences},
  month = Dec,
  pages = {A92},
}

@article{Wittenmyer2019,
  title = {Cool Jupiters greatly outnumber their toasty siblings: occurrence rates from the Anglo-Australian Planet Search},
  volume = {492},
  issn = {1365-2966},
  url = {http://dx.doi.org/10.1093/mnras/stz3436},
  doi = {10.1093/mnras/stz3436},
  number = {1},
  journal = {Monthly Notices of the Royal Astronomical Society},
  publisher = {Oxford University Press (OUP)},
  author = {Wittenmyer, Robert A and Wang, Songhu and Horner, Jonathan and Butler, R P and Tinney, C G and Carter, B D and Wright, D J and Jones, H R A and Bailey, J and O’Toole, S J and Johns, Daniel},
  year = {2019},
  month = Dec,
  pages = {377–383},
}

@article{Manara22,
  author = {{Manara}, C.~F. and {Ansdell}, M. and {Rosotti}, G.~P. and {Hughes}, A.~M. and {Armitage}, P.~J. and {Lodato}, G. and {Williams}, J.~P.},
  title = {{Demographics of young stars and their protoplanetary disks: lessons learned on disk evolution and its connection to planet formation}},
  journal = {arXiv e-prints},
  year = {2022},
  month = mar,
  eid = {arXiv:2203.09930},
  pages = {arXiv:2203.09930},
  doi = {10.48550/arXiv.2203.09930},
  archiveprefix = {arXiv},
  eprint = {2203.09930},
  primaryclass = {astro-ph.SR},
  adsurl = {https://ui.adsabs.harvard.edu/abs/2022arXiv220309930M},
}

@misc{VanZandt2026,
  title = {The {Giant} {Planet} {Mass} {Function} {Across} {Stellar} {Mass} and {Metallicity}: {Massive} {Stars} {Make} {Massive} {Planets}},
  shorttitle = {The {Giant} {Planet} {Mass} {Function} {Across} {Stellar} {Mass} and {Metallicity}},
  url = {https://scixplorer.org/abs/2026arXiv261103613V},
  doi = {10.48550/arXiv.2611.03613},
  urldate = {2026-11-21},
  publisher = {arXiv},
  author = {Van Zandt, Judah and Bowler, Brendan},
  month = nov,
  year = {2026},
  note = {ADS Bibcode: 2026arXiv261103613V},
}

@article{Pascucci2016,
  author = {{Pascucci}, I. and {Testi}, L. and {Herczeg}, G.~J. and {Long}, F. and {Manara}, C.~F. and {Hendler}, N. and {Mulders}, G.~D. and {Krijt}, S. and {Ciesla}, F. and {Henning}, Th. and {Mohanty}, S. and {Drabek-Maunder}, E. and {Apai}, D. and {Sz{\H{u}}cs}, L. and {Sacco}, G. and {Olofsson}, J.},
  title = {{A Steeper than Linear Disk Mass-Stellar Mass Scaling Relation}},
  journal = {\apj},
  year = {2016},
  month = nov,
  volume = {831},
  number = {2},
  eid = {125},
  pages = {125},
  doi = {10.3847/0004-637X/831/2/125},
  archiveprefix = {arXiv},
  eprint = {1608.03621},
  primaryclass = {astro-ph.EP},
  adsurl = {https://ui.adsabs.harvard.edu/abs/2016ApJ...831..125P},
}

@article{Ansdell2017,
  author = {{Ansdell}, M. and {Williams}, J.~P. and {Manara}, C.~F. and {Miotello}, A. and {Facchini}, S. and {van der Marel}, N. and {Testi}, L. and {van Dishoeck}, E.~F.},
  title = {{An ALMA Survey of Protoplanetary Disks in the {\ensuremath{\sigma}} Orionis Cluster}},
  journal = {\aj},
  year = {2017},
  month = may,
  volume = {153},
  number = {5},
  eid = {240},
  pages = {240},
  doi = {10.3847/1538-3881/aa69c0},
  archiveprefix = {arXiv},
  eprint = {1703.08546},
  primaryclass = {astro-ph.EP},
  adsurl = {https://ui.adsabs.harvard.edu/abs/2017AJ....153..240A},
}

@article{Bitsch2019,
  title = {Formation of planetary systems by pebble accretion and migration: growth of gas giants},
  volume = {623},
  issn = {1432-0746},
  url = {http://dx.doi.org/10.1051/0004-6361/201834489},
  doi = {10.1051/0004-6361/201834489},
  journal = {Astronomy \& Astrophysics},
  publisher = {EDP Sciences},
  author = {Bitsch, Bertram and Izidoro, Andre and Johansen, Anders and Raymond, Sean N. and Morbidelli, Alessandro and Lambrechts, Michiel and Jacobson, Seth A.},
  year = {2019},
  month = Mar,
  pages = {A88},
}

@article{Schlecker2021,
  title = {The New Generation Planetary Population Synthesis (NGPPS): III. Warm super-Earths and cold Jupiters: a weak occurrence correlation, but with a strong architecture-composition link},
  volume = {656},
  issn = {1432-0746},
  url = {http://dx.doi.org/10.1051/0004-6361/202038554},
  doi = {10.1051/0004-6361/202038554},
  journal = {Astronomy \& Astrophysics},
  publisher = {EDP Sciences},
  author = {Schlecker, M. and Mordasini, C. and Emsenhuber, A. and Klahr, H. and Henning, Th. and Burn, R. and Alibert, Y. and Benz, W.},
  year = {2021},
  month = Dec,
  pages = {A71},
}

@article{Santos2001,
  title = {The metal-rich nature of stars with planets},
  volume = {373},
  issn = {1432-0746},
  url = {http://dx.doi.org/10.1051/0004-6361:20010648},
  doi = {10.1051/0004-6361:20010648},
  number = {3},
  journal = {Astronomy \& Astrophysics},
  publisher = {EDP Sciences},
  author = {Santos, N. C. and Israelian, G. and Mayor, M.},
  year = {2001},
  month = July,
  pages = {1019–1031},
}
\bibliographystyle{aasjournalv7}



\end{document}